\documentclass[aps,onecolumn,pra,superscriptaddress]{revtex4-1}
\usepackage{amssymb}
\usepackage{amsmath}
\usepackage{graphicx}
\usepackage{subfigure}
\usepackage{natbib}
\usepackage{epsfig}
\usepackage{amsfonts}
\usepackage{mathrsfs}
\usepackage{xcolor}
\usepackage[toc,page,title,titletoc,header]{appendix}

\begin{document}

\title{Coherence enhanced quantum metrology in a nonequilibrium optical molecule}

\author{Zhihai \surname{Wang}}
\email{Z. Wang and W. Wu contributed equally to this paper.}
\affiliation{Center for Quantum Sciences and School of Physics, Northeast Normal
University, Changchun 130024, China\\
and Center for Advanced Optoelectronic Functional Materials Research, and Key
Laboratory for UV-Emitting Materials and Technology of Ministry of
Education, Northeast Normal University, Changchun 130024, China}
\affiliation{Beijing Computational Science Research Center, Beijing {100094}, China}

\author{Wei \surname{Wu}}
\email{Z. Wang and W. Wu contributed equally to this paper.}
\affiliation{State Key Laboratory of Electroanalytical Chemistry, Changchun Institute of Applied Chemistry, Chinese Academy of Sciences, Changchun 130022, China}

\author{Guodong \surname{Cui}}
\affiliation{Department of Physics and Astronomy, State University of New York at Stony Brook, NY 11794, USA}

\author{Jin \surname{Wang}}
\email{Email: jin.wang.1@stonybrook.edu}
\affiliation{State Key Laboratory of Electroanalytical Chemistry, Changchun Institute of Applied Chemistry, Chinese Academy of Sciences, Changchun 130022, China}
\affiliation{Department of Physics and Astronomy, State University of New York at Stony Brook, NY 11794, USA}
\affiliation{Department of Chemistry, State University of New York at Stony Brook, NY 11794, USA}

\begin{abstract}
We explore the quantum metrology in an optical molecular system coupled to two environments with different temperatures, using a quantum master equation beyond secular approximation. We discover that the steady-state coherence originating from and sustained by the nonequilibrium condition can enhance quantum metrology. We also study the quantitative measures of the nonequilibrium condition in terms of the curl flux, heat current and entropy production at the steady state. They are found to grow with temperature difference. However, an apparent paradox arises considering the contrary behaviors of the steady-state coherence and the nonequilibrium measures in relation to the inter-cavity coupling strength. This paradox is resolved by decomposing the heat current into a population part and a coherence part. Only the latter, coherence heat current, is tightly connected to the steady-state coherence and behaves similarly with respect to the inter-cavity coupling strength. Interestingly, the coherence heat current flows from the low-temperature reservoir to the high-temperature reservoir, opposite to the direction of the population heat current. Our work offers a viable way to enhance quantum metrology for open quantum systems through steady-state coherence sustained by the nonequilibrium condition, which can be controlled and manipulated to maximize its utility. The potential applications go beyond quantum metrology and extend to areas such as device designing, quantum computation and quantum technology in general.

\end{abstract}
\keywords{Coherence, Quantum Fisher information, heat current}
%\pacs{42.50.Pq, 03.67.Lx, 42.50.Dv}
%\date {\today}
%\preprint{PDC vesion2,Wang, 140627}
\maketitle

\section{Introduction}

As part of the emerging field of quantum technology~\cite{JL}, quantum metrology aims to make high precision measurements of physical parameters
by exploiting the quantum nature of a system. In recent years, there has been growing interest in the quantum metrology of various physical systems, including interacting spin systems~\cite{Marzolino1,Marzolino11,Marzolino2}, cold atoms~\cite{Marzolino5} and quantum gases~\cite{Marzolino6}, to name but a few. A central quantity in quantum metrology is the quantum Fisher information (QFI), the inverse of which, according to the quantum Cram\'er-Rao theorem, gives a lower bound on the variance of any unbiased estimator of a parameter~\cite{cave,cave1}. The use of QFI is not limited to quantum metrology, but also extends to other aspects of quantum physics, such as quantum cloning \cite{QuantumCloning1,QuantumCloning2}, entanglement detection \cite{EntanglementDetection1,EntanglementDetection2,EntanglementDetection3} and quantum phase transition \cite{PhaseTransition1,PhaseTransition2}. In principle, quantum resources, such as coherence and entanglement, may be employed to enhance quantum metrology beyond the classical shot-noise limit~\cite{QFIBeyondLimit1,QFIBeyondLimit2,QFIBeyondLimit3}. However, in practice, due to the inevitable interactions with the surrounding environment, most quantum systems quickly lose their quantum features through the decoherence process~\cite{br,wh}, thus limiting the efficiency and application of quantum metrology.

%The use of QFI is not limited to the field of quantum metrology. For instance, a recent study proposed that QFI can be used to detect the nonequilibrium phase transition~\cite{Marzolino3}.

Recently, it has been proposed that stable quantum features, such as steady-state coherence \cite{jin1, jin2,jin3, cpsun} and steady-state entanglement \cite{SSEntanglement,SSEntanglement2}, may exist in open quantum systems interacting with nonequilibrium environments that sustain quantum nonequilibrium steady states \cite{QuantumNESS1,QuantumNESS2,QuantumNESS4}. Such nonequilibrium environments can be bosonic or fermionic, with different temperatures and/or chemical potentials. The surviving quantum features are essentially sustained by the nonequilibrium condition (temperature difference and/or chemical potential difference) in the environments interacting with the quantum system \cite{jin1,jin2,jin3,cpsun,SSEntanglement,SSEntanglement2}. In a certain sense, these quantum features do not only survive, but also thrive, in the noisy nonequilibrium environments, as they are actually born out of the interactions with the nonequilibrium environments. In contrast to their overprotected counterparts in isolated quantum systems that wither away the moment they make contact with the outside world \cite{br,wh}, these steady-state quantum features grown up in the jungle of nonequilibrium environments are immune to decoherence \cite{jin1,jin2,jin3,cpsun,SSEntanglement,SSEntanglement2}. This makes them valuable assets to quantum metrology and quantum technology in general. In particular, steady-state coherence may be utilized to enhance QFI and thus benefit quantum metrology.

Another important characteristic of these open quantum systems in nonequilibrium environments is the presence of steady-state currents \cite{jin1, jin2,jin3, cpsun,QuantumNESS1,QuantumNESS2,QuantumNESS4}, associated with the continuous exchange of matter, energy or information between the system and the environments at the steady state. On the dynamical level, the steady-state current is manifested as a probability curl flux that signifies the breaking of detailed balance and time reversal symmetry at the steady state \cite{jin1, jin2,jin3}. On the thermodynamic level, it is represented by the heat current (or particle flow) at the steady state arising from the temperature difference (or chemical potential difference), which has been widely used to design thermal transport devices~\cite{TW}, such as thermal transistor~\cite{KJ}, diode~\cite{JO} and rectification~\cite{TW1,manz}. Connected to the heat current at the steady state is the entropy production rate (EPR) which serves as a quantifier of the amount of detailed balance breaking and time irreversibility \cite{jin1,jin2,jin3,man,nic,JS,XJ}. Moreover, it has been suggested that steady-state coherence and steady-state current may be closely related to each other \cite{jin1,cpsun}. Therefore, it is also important to investigate the quantitative connection between nonequilibrium transport processes associated with heat currents~\cite{Mu1,Mu2,NB} and quantum metrology enhanced by steady-state coherence.

In this paper, we investigate the above issues by studying an optical molecule composed of two linearly coupled degenerate single-mode cavities~\cite{mole1,mole2}, which is immersed in two reservoirs with different temperatures, each in contact with a cavity. The nonequilibrium nature of the environments is signified by the temperature difference of the two reservoirs. We show that a residual steady-state coherence emerges which is sustained by the nonequilibrium condition, by solving the quantum master equation (QME) beyond secular approximation at the steady state. We find that the steady-state coherence augments the QFI and can effectively enhance the quantum metrology when the inter-cavity coupling is not too strong. Furthermore, we quantify the nonequilibrium measures in terms of the curl flux, heat current and EPR. These nonequilibrium measures are found to grow with temperature difference as anticipated. However, a paradox seems to emerge as the steady-state coherence and the nonequilibrium measures display opposite trends with respect to the inter-cavity coupling strength. We resolve this paradox by showing that the heat current can be decomposed into a population component and a coherence component. Only the latter part of the heat current is closely tied to the steady-state coherence and shows similar behaviors in relation to the inter-cavity coupling strength. Curiously, we find this coherence heat current flows from the low-temperature reservoir to the high-temperature reservoir, but does not violate the second law of thermodynamics when the population heat current is also considered.

The rest of the paper is organized as follows. In Sec.~\ref{model}, we present the model studied and derive the QME beyond secular approximation. In Sec.~\ref{coherence}, we solve the steady state of the QME and show that steady-state coherence can be used to enhance quantum metrology. In Sec.~\ref{none}, we quantify the nonequilibrium measures in terms of the curl flux, heat current and EPR. In Sec.~\ref{explain}, we resolve the apparent paradox through the decomposition of the heat current. Finally, some remarks on experimental realization and conclusion are given in Sec.~\ref{conclusion}.

\section{Model and QME}
\label{model}
\begin{figure}[tbp]
\centering
\includegraphics[width=8cm]{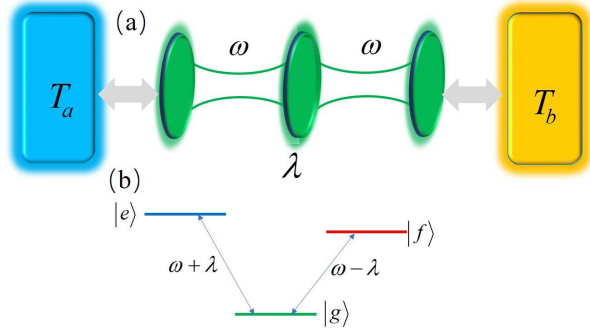}
\caption{(Color online) (a) The schematic representation of an optical molecule consisting of two coupled identical single-mode cavities, each interacting with its own reservoir with a different temperature ($T_a<T_b$).(b) The energy-level diagram in the supermode representation.}
\label{scheme}
\end{figure}

The optical-molecular system under consideration is schematically shown in Fig.~\ref{scheme}(a). The two coupled identical single-mode cavities are immersed, respectively, in their own reservoirs with different temperatures. The total Hamiltonian of the system plus reservoirs reads $H=H_{s}+H_{B}+H_{I}$, where (natural unit $\hbar=1$ is used throughout)
\begin{subequations}
\begin{eqnarray}
H_s&=&\omega(a^{\dagger}a+b^{\dagger}b)+\lambda(a^{\dagger}b+ab^{\dagger}),\\
H_{B}&=&\sum_{k}\omega_{ck}c_{k}^{\dagger}c_{k}+\sum_{k}\omega_{dk}d_{k}^{\dagger}d_{k},\\
H_{I}&=&\sum_{k}g_{k}(a^{\dagger}c_{k}+ac_{k}^{\dagger})+\sum_{k}f_{k}(b^{\dagger}d_{k}+bd_{k}^{\dagger}).
\end{eqnarray}
\end{subequations}

Here $H_s$ is the Hamiltonian for the optical molecule, describing two degenerately coupled cavity modes with annihilation operators $a$ and $b$, respectively. The parameter $\omega$ is the resonant frequency and $\lambda$ is the inter-cavity coupling strength. $H_B$ is the free Hamiltonian of the reservoirs, where $c_k(d_k)$ is the annihilation operator of the $k$th bosonic reservoir mode in contact with cavity $a(b)$, and $\omega_{ck(dk)}$ is the corresponding frequency. $H_I$ is the interaction Hamiltonian between the optical molecule and the reservoirs, where $g_k$ and $f_k$ are the coupling strengths between the $k$th mode in the reservoirs and the cavity mode $a$ and $b$, respectively. We assume that $\lambda,g_k$ and $f_k$ are all real.

The Hamiltonian $H_s$ can be diagonalized as $H_{s}=\omega_{A}A^{\dagger}A+\omega_{B}B^{\dagger}B$, by introducing the global operators $A=(a+b)/\sqrt{2}$ and $B=(a-b)/\sqrt{2}$ that represent two supermodes, where $\omega_{A}=\omega+\lambda$ and $\omega_{B}=\omega-\lambda$.  In terms of the operators $A$ and $B$, the system-reservoir interaction Hamiltonian can be reformulated as
\begin{equation}
H_{I}=\frac{1}{\sqrt{2}}\sum_{k}g_{k}[(A^{\dagger}+B^{\dagger})c_{k}+(A+B)c_{k}^{\dagger}]
+\frac{1}{\sqrt{2}}\sum_{k}f_{k}[(A^{\dagger}-B^{\dagger})d_{k}+(A-B)d_{k}^{\dagger}].
\end{equation}
In the interaction picture with the free Hamiltonian $H_0=H_s+H_B$, we have
\begin{equation}
H_{I}(t) = V_{C}^{-}(t)+V_{C}^{+}(t)+V_{D}^{-}(t)+V_{D}^{+}(t),
\end{equation}
where $V_{C}^{-}(t)=[V_{C}^{+}(t)]^{\dagger}=\frac{1}{\sqrt{2}}\sum_{k}g_{k}(A^{\dagger}e^{i\omega_{A}t}
+B^{\dagger}e^{i\omega_{B}t})c_{k}e^{-i\omega_{ck}t}$, $V_{D}^{-}(t)=[V_{D}^{+}(t)]^{\dagger}
=\frac{1}{\sqrt{2}}\sum_{k}f_{k}(A^{\dagger}e^{i\omega_{A}t}
-B^{\dagger}e^{i\omega_{B}t})d_{k}e^{-i\omega_{dk}t}$.

Under the Born-Markov approximation, the QME in the interaction picture reads \cite{br}
\begin{equation}
\frac{d\rho_I}{dt}=-\int_{0}^{\infty}ds{\rm Tr}_{B}[H_{I}(t),[H_{I}(t-s),\rho_I(t)\otimes\rho_B]],
\end{equation}
where $\rho_I$ is the reduced density operator of the system in the interaction picture, $\rho_B$ is the density operator of the reservoirs (each reservoir remains at its thermal equilibrium state under the Born approximation), and $\rm{Tr}_B$ denotes the partial trace with respect to the degrees of freedom of the reservoirs.

Going back to the Schr\"odinger picture, \emph{without} making the secular approximation, we finally arrive at the QME for the reduced density operator of the system
\begin{equation}
\frac{d\rho}{dt}=-i[H_s,\rho]+D_0[\rho]+D_s[\rho],
\end{equation}
where
\begin{eqnarray}
D_{0}[\rho]&=&\eta_{+}^{A}(2A^{\dagger}\rho A-\rho AA^{\dagger}-AA^{\dagger}\rho)+\eta_{+}^{B}(2B^{\dagger}\rho B-\rho BB^{\dagger}-BB^{\dagger}\rho)\nonumber\\
&+&\xi_{+}^{A}(2A\rho A^{\dagger}-\rho A^{\dagger}A-A^{\dagger}A\rho)+\xi_{+}^{B}(2B\rho B^{\dagger}-\rho B^{\dagger}B-B^{\dagger}B\rho),\label{D0}
\end{eqnarray}
and
\begin{eqnarray}
D_{s}[\rho]&=&(\eta_{-}^{A}+\eta_{-}^{B})A^{\dagger}\rho B
-\eta_{-}^{A}BA^{\dagger}\rho-\eta_{-}^{B}\rho BA^{\dagger} +h.c.\nonumber\\
&+&(\xi_{-}^{A}+\xi_{-}^{B})B\rho A^{\dagger}
-\xi_{-}^{A}\rho A^{\dagger}B-\xi_{-}^{B}A^{\dagger}B\rho+h.c..\label{D1}
\end{eqnarray}

In the above, we have defined $\eta_{\pm}^{A (B)}:=\eta_{\pm}(\omega_{A(B)})$ and $\xi_{\pm}^{A(B)}:=\xi_{\pm}(\omega_{A(B)})$, where
\begin{eqnarray}
\eta_{\pm}(\omega):=\frac{1}{2}[\gamma_{a}(\omega)N_{a}(\omega)\pm \gamma_{b}(\omega)N_{b}(\omega)],\quad \xi_{\pm}(\omega):=\frac{1}{2}\{\gamma_{a}(\omega)[N_{a}(\omega)+1]\pm \gamma_{b}(\omega)[N_{b}(\omega)+1]\}.
\label{parameter}
\end{eqnarray}
Here $\gamma_{a}(\omega)=\pi\sum_{k}g_{k}^{2}\delta(\omega-\omega_{ck})$ and $\gamma_{b}(\omega)
=\pi\sum_{k}f_{k}^{2}\delta(\omega-\omega_{dk})$ are the spectral densities of the reservoirs in contact with cavity $a$ and $b$, respectively. $N_\alpha(\omega):=[\exp(\omega/T_\alpha)-1]^{-1}$($\alpha=a,b$ and $k_B=1$ in natural units) is the Planck distribution for the reservoirs, describing the average Bose occupation number on frequency $\omega$  at temperature $T_{\alpha}$ in the reservoir. For simplicity, we shall assume that the spectra of the reservoirs are frequency independent and restrict ourselves to the balanced coupling regime, that is, $\gamma_a(\omega)=\gamma_b(\omega)=\gamma$. As a result, $\eta_{+}(\omega)=\gamma N_{+}(\omega)$, $\xi_{+}(\omega)=\gamma[N_{+}(\omega)+1]$, and $\eta_{-}(\omega)=\xi_{-}(\omega)=\gamma N_{-}(\omega)$, where $N_{\pm}(\omega):=[N_{a}(\omega)\pm N_{b}(\omega)]/2$. We also introduce the short notations $N_{\pm}^{A (B)}:=N_{\pm}(\omega_{A(B)})$, so that $\eta_{+}^{A(B)}=\gamma N_{+}^{A(B)}$, $\xi_{+}^{A(B)}=\gamma(N_{+}^{A(B)}+1)$, and $\eta_{-}^{A(B)}=\xi_{-}^{A(B)}=\gamma N_{-}^{A(B)}$.

The dissipation terms in Eqs.~(\ref{D0}) and (\ref{D1}) represent second order processes. The dissipator $D_0[\rho]$ describes the process of the supermode $A$ or $B$ absorbing an energy quantum in the reservoirs, and emitting it back to the reservoirs by the same supermode. The dissipator $D_s[\rho]$, on the other hand, is associated with the absorption and emission of the energy quantum completed by different supermodes. Since the two supermodes are carrying different frequencies, i.e., $\omega_A\neq\omega_B$, the dissipator $D_s[\rho]$ is often considered as terms with high frequency and thus neglected by performing the so-called ``secular approximation"~\cite{TW,KJ,JO,TW1}. In the equilibrium situation $T_a=T_b$, the secular approximation does give a reasonable result. More specifically, for balanced coupling ($\gamma_a(\omega)=\gamma_b(\omega)=\gamma$), the equilibrium condition $T_a=T_b$ yields exactly vanishing $D_s[\rho]$. This is because, according to Eq. (\ref{parameter}), parameters with a negative subscript ($\eta_{-}^{A(B)}$ and $\xi_{-}^{A(B)}$ as well as $N_{-}^{A(B)}$) vanish exactly at equilibrium with $T_a=T_b$.  However, in nonequilibrium environments for reservoirs with different temperatures ($T_a\neq T_b$), the secular approximation will disregard the dissipator $D_s[\rho]$ that can induce important quantum effects, such as steady-state coherence~\cite{jin1, jin2,jin3, cpsun}, as will be shown in a moment. Therefore, in our treatment the dissipator $D_s[\rho]$ is retained without performing the secular approximation.

\section{Quantum metrology enhanced by steady-state coherence}\label{coherence}

\subsection{Steady State of the QME}
We have obtained the QME beyond the secular approximation by taking the dissipator $D_s[\rho]$ into account. A direct consequence is the presence of steady-state quantum coherence in the nonequilibrium regime. For simplicity, we restrict ourselves to the subspace of zero and single photon excitations, which is reasonable at low temperatures. This subspace is spanned by three basis vectors $\{|g\rangle:=|0,0\rangle,|e\rangle:=|1,0\rangle,|f\rangle:=|0,1\rangle\}$, where $|m,n\rangle:=|m\rangle_A\otimes|n\rangle_B$. In this subspace, we have $A=|g\rangle\langle e|$, $B=|g\rangle\langle f|$, and $H_s=\omega_A |e\rangle\langle e|+\omega_B |f\rangle\langle f|$. The energy-level diagram is sketched in Fig.~\ref{scheme}(b). The QME for the density matrix elements in the subspace reads
\begin{subequations}
\begin{eqnarray}
\frac{d}{dt}\rho_{gg} & = & -2\gamma(N_{+}^{A}+N_{+}^{B})\rho_{gg}+2\gamma(N_{+}^{A}+1)\rho_{ee}+2\gamma(N_{+}^{B}+1)\rho_{ff}
+\gamma(N_{-}^{A}+N_{-}^{B})(\rho_{fe}^{\rm coh}+\rho_{ef}^{\rm coh}),\label{eq:gg}\\
\frac{d}{dt}\rho_{ee} & = & 2\gamma N_{+}^{A} \rho_{gg}-2\gamma(N_{+}^{A}+1)\rho_{ee}-\gamma N_{-}^{B}(\rho_{fe}^{\rm coh}+\rho_{ef}^{\rm coh}),\label{eq:ee}\\
\frac{d}{dt}\rho_{ff} & = & 2\gamma N_{+}^{B}\rho_{gg}-2\gamma(N_{+}^{B}+1)\rho_{ff}-\gamma N_{-}^{A}(\rho_{fe}^{\rm coh}+\rho_{ef}^{\rm coh}),\label{eq:ff}\\
\frac{d}{dt}\rho_{ef}^{\rm coh} & = & \gamma(N_{-}^{A}+N_{-}^{B})\rho_{gg}-\gamma N_{-}^{A}\rho_{ee}-\gamma N_{-}^{B}\rho_{ff}-\left[i(\omega_{A}-\omega_{B})+\gamma(N_{+}^{A}+N_{+}^{B}+2)\right]\rho_{ef}^{\rm coh},\label{eq:offd}\\
\frac{d}{dt}\rho_{ge}^{\rm coh} & = & \left[i\omega_{A}-\gamma(2N_{+}^{A}+N_{+}^{B}+1)\right]\rho_{ge}^{\rm coh}-\gamma N_{-}^{B}\rho_{gf}^{\rm coh},\\
\frac{d}{dt}\rho_{gf}^{\rm coh} & = & \left[i\omega_{B}-\gamma(2N_{+}^{B}+N_{+}^{A}+1)\right]\rho_{gf}^{\rm coh}-\gamma N_{-}^{A}\rho_{ge}^{\rm coh}.
\end{eqnarray}\label{elements}
\end{subequations}
The superscript ``$\rm{coh}$'' has been used to indicate coherence of the quantum system represented by the off-diagonal elements of the density matrix.

We notice that, in the above set of equations, two coherence variables, $\rho_{ge}^{\rm{coh}}$ and $\rho_{gf}^{\rm{coh}}$, are decoupled from the rest of the variables.  However, the coherence variable $\rho^{\rm{coh}}_{ef}$ is coupled, in the nonequilibrium regime, to the populations in the density matrix, $\rho_{ii}\,(i=g,e,f)$, as a result of  the dissipator $D_s[\rho]$. Consequently, in the long time limit when the steady state is reached, the steady-state density matrix has the form
\begin{equation}
\rho_{ss}=\left(\begin{array}{ccc}
\rho_{gg}^{ss} & 0 & 0\\
0 & \rho_{ee}^{ss} & \rho^{\rm{coh}}_{ef}\\
0 & \rho^{\rm{coh}*}_{ef} & \rho_{ff}^{ss}
\end{array}\right),
\label{ss}
\end{equation}
where the superscript ``$ss$" has been used to indicate steady-state populations, but not for the steady-state coherence $\rho^{\rm{coh}}_{ef}$ as that will complicate the notation too much. The steady-state coherence $\rho^{\rm{coh}}_{ef}$ is in general nonvanishing in the nonequilibrium regime ($T_a\neq T_b$) due to its coupling with the populations that arises from the dissipator $D_s[\rho]$.

Moreover, we have obtained the analytical expressions of the steady-state density matrix elements. (The method of obtaining this analytical solution is outlined in Appendix \ref{SolutionMethod}.) The steady-state populations are given by
\begin{equation}\label{Population2}
\left\{
\begin{array}{rcl}
\rho_{gg}^{ss}&=&\left[(N_{+}^{A}+1)(N_{+}^{B}+1)-(N_{+}^{A}+N_{+}^{B}+2)N_{-}^{A}N_{-}^{B}R\right]/\mathcal{N}\\
&&\\
\rho_{ee}^{ss}&=&\left[N_{+}^{A}(N_{+}^{B}+1)-(N_{+}^{A}+N_{+}^{B}+1)N_{-}^{A}N_{-}^{B}R-(N_{-}^{B})^2R\right]/\mathcal{N}\\
&&\\
\rho_{ff}^{ss}&=&\left[N_{+}^{B}(N_{+}^{A}+1)-(N_{+}^{A}+N_{+}^{B}+1)N_{-}^{A}N_{-}^{B}R-(N_{-}^{A})^2R\right]/\mathcal{N}\\
\end{array}
\right.,
\end{equation}
and the steady-state coherence reads
\begin{equation}\label{SSCES2}
\rho_{ef}^{\rm{coh}}=\frac{N_{-}^{A}(N_{+}^{B}+1)+N_{-}^{B}(N_{+}^{A}+1)}{ \mathcal{N}[(N_{+}^{A}+N_{+}^{B}+2)+i(\omega_A-\omega_B)/\gamma]}.
\end{equation}
In the above, $R$ has the expression
\begin{equation}\label{ExpR}
R=\frac{(N_{+}^{A}+N_{+}^{B}+2)}{(N_{+}^{A}+N_{+}^{B}+2)^2+[(\omega_A-\omega_B)/\gamma]^2}
\end{equation}
and $\mathcal{N}$ is a normalization factor fixed by the condition $\rho_{gg}^{ss}+\rho_{ee}^{ss}+\rho_{ff}^{ss}=1$ (explicit expression given in Appendix \ref{Expressions}).

It is instructive to examine the steady-state solution at equilibrium with $T_a=T_b=T$. Notice that at equilibrium we have $N_{-}^{A}=N_{-}^{B}=0$ and $N_{+}^{A(B)}=N_{a}^{A(B)}=N_{b}^{A(B)}:=N^{A(B)}$. Immediately, we see from Eq. (\ref{SSCES2}) that the steady-state coherence vanishes, i.e., $\rho_{ef}^{\rm{coh}}=0$ at equilibrium, due to the vanishing factors $N_{-}^{A}$ and $N_{-}^{B}$ at equilibrium. Furthermore, Eq. (\ref{Population2}) shows that the equilibrium populations are given by $\rho_{gg}^{eq}=(N^{A}+1)(N^{B}+1)/\mathcal{N}$, $\rho_{ee}^{eq}=N^{A}(N^{B}+1)/\mathcal{N}$, $\rho_{ff}^{eq}=N^{B}(N^{A}+1)/\mathcal{N}$, where the subscript ``$eq$" indicates equilibrium. Alternatively, we have $\rho_{ee}^{eq}/\rho_{gg}^{eq}=N^A/(N^A+1)=e^{-\omega_A/T}$ and $\rho_{ff}^{eq}/\rho_{gg}^{eq}=N^B/(N^B+1)=e^{-\omega_B/T}$. These results agree with the thermal-state density operator $\rho_{eq}=Z^{-1} \exp\{-H_s/T\}$ for the system in contact with an equilibrium reservoir.

\begin{figure}[tbp]
%\centering
\includegraphics[width=0.8\columnwidth]{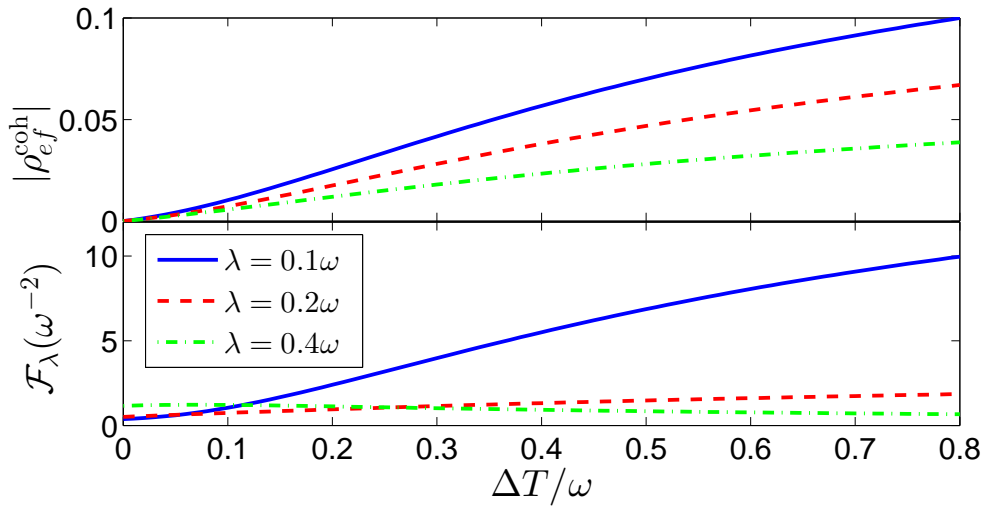}
\caption{(Color online)  (a) The steady-state coherence $|\rho^{\rm{coh}}_{ef}|$ as a function of temperature difference $\Delta T$. (b) The QFI $\mathcal{F}_\lambda$ as a function of temperature difference $\Delta T$. The parameters are set as $\gamma=0.1\omega$, $T_a=0.2\omega$.}
\label{coherencefisher}
\end{figure}

In the nonequilibrium regime with $T_a\neq T_b$, the steady-state coherence $\rho^{\rm{coh}}_{ef}$ is in general nonvanishing, stable against decoherence as it is essentially sustained by the nonequilibrium condition. The nonequilibrium condition indicated by $T_a\neq T_b$ is manifested in the steady-state coherence through the nonvanishing factors $N_{-}^{A}$ and $N_{-}^{B}$ given that $N_{-}^{A(B)}=(N_{a}^{A(B)} - N_{b}^{A(B)})/2=\{[\exp(\omega_{A(B)}/T_a)-1]^{-1} - [\exp(\omega_{A(B)}/T_b)-1]^{-1}\}/2$. (Note that $N_{-}^{A}$ and $N_{-}^{B}$ are also implicit in the normalization factor $\mathcal{N}$.) We could also use  the temperature difference $\Delta T:=|T_b-T_a|$ to characterize the strength of the nonequilibrium condition (i.e., the degree of `nonequilibriumness'). The magnitude of the steady-state coherence can be quantified by $|\rho^{\rm{coh}}_{ef}|$ (the modulus of $\rho^{\rm{coh}}_{ef}$). Considering that the magnitude of $N_{-}^{A (B)}$ increases with the temperature difference, Eq. (\ref{SSCES2}) suggests that the magnitude of the steady-state coherence also grows with the temperature difference at least when $\Delta T$ is not too large. In Fig.~\ref{coherencefisher}(a), $|\rho^{\rm{coh}}_{ef}|$ is plotted as a function of the temperature difference $\Delta T$ (with $T_a$ fixed), for different inter-cavity coupling strength $\lambda$. As can be seen, the steady-state coherence increases with $\Delta T$ for fixed $\lambda$, in the parameter regime considered. One can also see that, for fixed $\Delta T$, the steady-state coherence decreases with the inter-cavity coupling strength $\lambda$. Since $\lambda$ characterizes the energy-splitting between the supermodes [as shown in Fig.~\ref{scheme}(b)], a larger $\lambda$ naturally corresponds to a smaller coherence.

\subsection{Quantum Fisher Information}

As a potential application of the steady-state coherence sustained by the nonequilibrium condition, we investigate how it can assist with quantum metrology characterized by QFI. The inverse of QFI gives a lower bound on the variance of any unbiased estimator of a physical parameter $\theta$~\cite{cave,cave1}. For a general quantum state described by the density matrix $\rho(\theta)$, with the spectral decomposition $\rho=\sum_{i}^{M}p_i|\psi_i\rangle \langle\psi_i|$ (where $M$ denotes the number of nonzero $p_i$), the QFI is given by~\cite{jing,wang1,wang2}
\begin{equation}
\mathcal{F}_{\theta}=\sum_{i=1}^{M}\frac{(\partial_{\theta}p_{i})^{2}}{p_{i}}+
4\sum_{i=1}^{M}p_{i}\langle\partial_{\theta}\psi_{i}|\partial_{\theta}\psi_{i}
\rangle-\sum_{i,j=1}^{M}\frac{8p_{i}p_{j}}{p_{i}+p_{j}}|\langle\psi_{i}|
\partial_{\theta}\psi_{j}\rangle|^{2}.
\label{commonfisher}
\end{equation}
A point probably worth mentioning is that a parameter-dependent phase change in the eigen-state $|\psi_i'\rangle=e^{i f_i(\theta)}|\psi_i\rangle$ would not alter the end result of $\mathcal{F}_{\theta}$ in Eq. (\ref{commonfisher}) considering that the eigen-states are orthonormal.

%This agrees with the fact that QFI is representation-independent.

For our system, direct calculation gives the spectral decomposition of the steady-state density matrix in Eq.~(\ref{ss}), with the eigen-values and eigen-vectors given by
\begin{eqnarray}
p_{1} & = & \rho_{gg}^{ss},\quad |\psi_{1}\rangle=|g\rangle,\\
p_{2} & = & (\rho_{ee}^{ss}+\rho_{ff}^{ss})/2+\sqrt{(\rho_{ee}^{ss}-\rho_{ff}^{ss})^{2}/4+|\rho_{ef}^{\rm coh}|^{2}},\quad |\psi_{2}\rangle=\cos\frac{\alpha}{2}e^{i\phi}|e\rangle+\sin\frac{\alpha}{2}|f\rangle,\\
p_{3} & = & (\rho_{ee}^{ss}+\rho_{ff}^{ss})/2-\sqrt{(\rho_{ee}^{ss}-\rho_{ff}^{ss})^{2}/4+|\rho_{ef}^{\rm coh}|^{2}}, \quad |\psi_{3}\rangle=\sin\frac{\alpha}{2}e^{i\phi}|e\rangle-\cos\frac{\alpha}{2}|f\rangle,
\end{eqnarray}
where
\begin{equation}\label{BetaPhiExpression}
\alpha=\arctan[2|\rho^{\rm{coh}}_{ef}|/(\rho_{ee}^{ss}-\rho_{ff}^{ss})], \quad \phi=\arg(\rho^{\rm{coh}}_{ef}).
\end{equation}
Notice that $\alpha$ and $\phi$ are both dependent on the steady-state coherence which vanishes at equilibrium.

Taking the inter-cavity coupling strength $\lambda$ as the estimated physical parameter, we obtain, according to Eq. (\ref{commonfisher}), the QFI with the following expression
\begin{equation}
\mathcal{F}_{\lambda}=\sum_{i=1}^{3}\frac{(\partial_{\lambda}p_{i})^{2}}{p_{i}}
+\frac{(p_2-p_3)^2}{p_2+p_3}\left[(\partial_{\lambda}\alpha)^{2}+(\partial_{\lambda}\phi)^{2}\sin^2\alpha\right].
\label{QFI}
\end{equation}
The first term in Eq.~(\ref{QFI}) represents the classical part of the QFI, which is contributed only by the diagonal elements of the density matrix (in its spectral decomposition representation). The second term results from the contribution of the steady-state quantum coherence (through $\alpha$ and $\phi$) sustained by the nonequilibrium condition. This term vanishes at equilibrium since vanishing steady-state coherence at equilibrium ($\rho^{\rm{coh}}_{ef}=0$) leads to $\alpha=0$ according to Eq. (\ref{BetaPhiExpression}), resulting in $\partial_{\lambda}\alpha=0$ and $\sin \alpha=0$ in Eq. (\ref{QFI}). Notice that this second term arising from the steady-state coherence is non-negative (typically positive under nonequilibrium conditions). It implies that under nonequilibrium conditions the steady-state coherence in general makes a positive contribution to (i.e., increases) the QFI on top of its classical part. Therefore, one could expect a close connection (not necessarily a simple one though) between the steady-state coherence and the QFI in relation to the nonequilibrium condition and the physical parameter estimated (the inter-cavity coupling strength in this case).

In Fig.~\ref{coherencefisher}(b), $\mathcal{F}_\lambda$ is plotted as a function of $\Delta T$ for different $\lambda$. We can see that the temperature difference measuring the degree of nonequilibriumness is able to enhance the QFI under certain conditions. This enhancement effect is especially prominent when the inter-cavity coupling is weak (e.g., $\lambda=0.1\omega$, blue line). In this weak inter-cavity coupling regime, the QFI behaves qualitatively similar to the steady-state coherence, monotonically increasing with the temperature difference. As the inter-cavity coupling grows stronger (e.g., $\lambda=0.2\omega$, red line), this enhancement effect, however, becomes less prominent, even though the QFI still increases monotonically with the temperature difference in the parameter regime considered. For $\lambda=0.4\omega$ (green line), the QFI almost stays constant as $\Delta T$ is increased. Numerical values indicate that it actually increases a little first and then decreases slowly as $\Delta T$ is further increased, displaying a non-monotonic behavior in the parameter regime. This reflects the complexity of the interplay between the QFI and the steady-state coherence in relation to the nonequilibrium condition and the physical parameter estimated.

Taken together, what we can say in this case is that, in the relatively weak inter-cavity coupling regime, the steady-state coherence sustained by the nonequilibrium condition is an effective booster of the QFI, capable of enhancing quantum metrology with a higher precision of parameter estimation. Thus one can manipulate the nonequilibrium condition to effectively augment quantum metrology through the steady-state coherence in the weak inter-cavity coupling regime. It is worth noting that, from a practical perspective of parameter estimation, weak inter-cavity coupling is probably also the most relevant regime, since its value is typically much more difficult to estimate due to its weakness. Our results indicate that, particularly in this regime, manipulating the nonequilibrium condition is capable of improving the precision of its estimation through QFI enhanced by steady-state coherence. In the next section we investigate the nonequilibrium measures of the system in terms of the curl flux, heat current and EPR at the steady state.

\section{Curl flux, heat current and EPR}\label{none}

\subsection{Circulating Curl Flux}
The QME can also be reformulated in a vector-matrix form, $|\dot\rho\rangle=\mathcal{M}|\rho\rangle$, by writing the elements of the density matrix as a vector and the dynamical generator as a matrix. More specifically, the vector $|\rho\rangle$ has the form $|\rho\rangle=(\rho_p, \rho_c)^{T}$, where $\rho_p$ is a vector representing the population component (diagonal elements of the density matrix) and $\rho_c$ is a vector representing the coherence component (off-diagonal elements of the density matrix). Then the dynamical generator $\mathcal{M}$ takes on a block matrix form, and the QME has the following form
\begin{equation}\label{LSQME}
\left(\begin{array}{c}
\dot{\rho}_{p}\\
\dot{\rho}_{c}
\end{array}\right)=\left(\begin{array}{cc}
\mathcal{M}_{p} & \mathcal{M}_{pc}\\
\mathcal{M}_{cp} & \mathcal{M}_{c}
\end{array}\right)\left(\begin{array}{c}
\rho_{p}\\
\rho_{c}
\end{array}\right).
\end{equation}
For our particular system, in the basis of the energy eigenstates $\{|g\rangle,|e\rangle,|f\rangle\}$, we have the population component $\rho_p=(\rho_{gg},\rho_{ee},\rho_{ff})^{T}$ and the coherence component $\rho_c=(\rho_{ef}^{\rm{coh}},\rho_{fe}^{\rm{coh}})^{T}$. (Note that in the coherence component we have excluded the other two coherence elements $\rho_{ge}^{\rm{coh}}$ and $\rho_{gf}^{\rm{coh}}$ as well as their complex conjugates, since their dynamics is decoupled from the rest according to Eq. (\ref{elements}) and is of no particular interest.) The matrix $\mathcal{M}$ as well its four blocks can be directly read off from Eq. (\ref{elements}).

Under suitable conditions (namely, all eigenvalues of $\mathcal{M}$ have negative real parts except one simple zero eigenvalue), the QME has a unique steady state $|\rho_{ss}\rangle$ that will be reached in the long time limit. At the steady state, we can eliminate the coherence component $\rho_c^{ss}$ to arrive at a steady-state equation for the population component only \cite{jin1,jin2,jin3}
\begin{equation}
(\mathcal{M}_p-\mathcal{M}_{pc}\mathcal{M}_c^{-1}\mathcal{M}_{cp})\rho^{ss}_{p}=0.
\label{landscale}
\end{equation}
Formally, it resembles a classical master equation for the steady state. One can introduce a transfer matrix $\mathcal{T}$ defined as $\mathcal{T}_{mn}=0$ for $m=n$ and $\mathcal{T}_{mn}=\mathcal{A}_{nm}\rho_{p; m}^{ss}$ for $m\neq n$, where $\mathcal{A}=\mathcal{M}_p-\mathcal{M}_{pc}\mathcal{M}_c^{-1}\mathcal{M}_{cp}$. The expressions of the matrix elements of $\mathcal{A}$ for our particular system are given in Appendix \ref{SolutionMethod}.

The transfer matrix $\mathcal{T}$ associated with the population dynamics can be further decomposed into two parts with different meanings ~\cite{jin1,jin2,jin3}.  For our system this decomposition reads
\begin{equation}\label{DTM}
\mathcal{T}=\left(\begin{array}{ccc}
0&\mathcal{A}_{eg}\rho^{ss}_{gg}&\mathcal{A}_{gf}\rho^{ss}_{ff}\\
\mathcal{A}_{eg}\rho^{ss}_{gg}&0&\mathcal{A}_{fe}\rho^{ss}_{ee}\\
\mathcal{A}_{gf}\rho^{ss}_{ff}&\mathcal{A}_{fe}\rho^{ss}_{ee}&0
\end{array}\right)+\mathcal{J}_{\rm curl}\left(\begin{array}{ccc}
0&0&1\\1&0&0\\
0&1&0
\end{array}\right),
\end{equation}
where
\begin{equation}\label{Jcurl}
\mathcal{J}_{\rm curl}=\mathcal{A}_{ge}\rho_{ee}^{ss}-\mathcal{A}_{eg}\rho_{gg}^{ss}=\mathcal{A}_{fg}\rho_{gg}^{ss}-\mathcal{A}_{gf}\rho_{ff}^{ss}= \mathcal{A}_{ef}\rho_{ff}^{ss}-\mathcal{A}_{fe}\rho_{ee}^{ss}.
\end{equation}
The equivalence of these expressions for $\mathcal{J}_{\rm curl}$ is guaranteed by the steady-state equation for the population component, Eq. (\ref{landscale}), and the property that the column elements of the matrix $\mathcal{A}$ sum to zero resulting from probability conservation (this property can be verified from the expression of $\mathcal{A}$ in Appendix \ref{Expressions}).

\begin{figure}[tbp]
%\centering
\includegraphics[width=0.8\columnwidth]{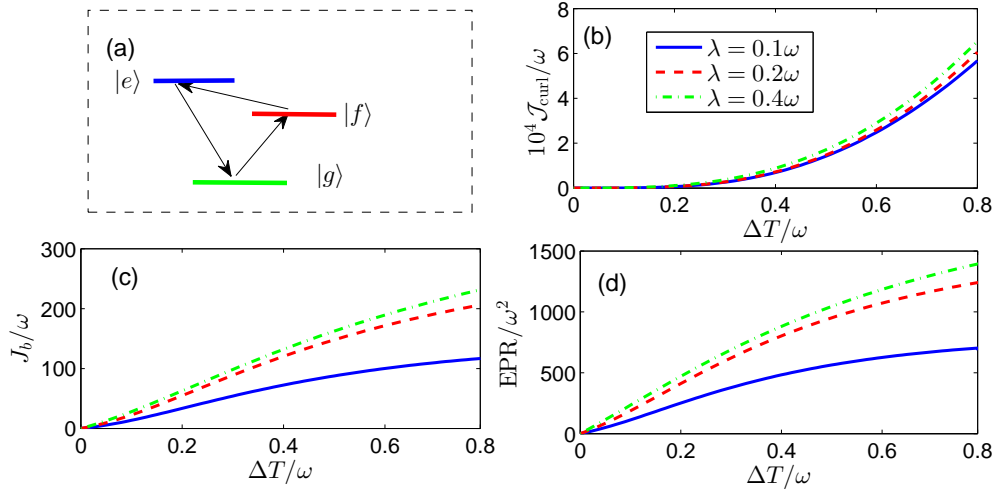}
\caption{(Color online) (a) A schematic representation of the curl flux, with the arrows indicating the direction of its circulation. (b) The curl flux $\mathcal{J}_{\rm curl}$ as a function of temperature difference $\Delta T$. (c) The heat current $J_b$ as a function of temperature difference $\Delta T$. (d) The EPR as a function of temperature difference $\Delta T$. The parameters are set as $\gamma=0.1\omega$, $T_a=0.2\omega$.}
\label{flux}
\end{figure}

The first part of the transfer matrix in Eq. (\ref{DTM}) is associated with the equilibrium reversal dynamics driven by the steady-state population landscape that satisfies the detailed balance condition indicating time reversibility. The second part is associated with the nonequilibrium irreversible dynamics driven by the curl flux circulating in a loop [see Fig.~\ref{flux}(a) for an illustration] that breaks detailed balance and time reversal symmetry. The nonequilibrium population dynamics is thus driven by both the steady-state population landscape and the circulating curl flux.

The circulating nature of the curl flux $\mathcal{J}_{\rm curl}$ is clear from the equivalence of its three expressions in Eq. (\ref{Jcurl}). Its connection to the nonequilibrium condition can be seen as follows.  Noticing that $\mathcal{A}_{ef}$ and $\mathcal{A}_{fe}$ in the last expression of $\mathcal{J}_{\rm curl}$ in Eq. (\ref{Jcurl}) have simple forms (see Appendix \ref{Expressions}), we obtain the following expression for the curl flux
\begin{equation}
\mathcal{J}_{\rm curl}=2\gamma R \left[(N_-^B)^2 \rho_{ff}^{ss}-(N_-^A)^2 \rho_{ee}^{ss}\right].\\
\end{equation}
In this expression $N_-^A$ and $N_-^B$ are directly tied to the nonequilibrium condition $T_{a}\neq T_{b}$ since they vanish at equilibrium. Thus we see immediately that the curl flux $\mathcal{J}_{\rm curl}$ also vanishes at equilibrium when $T_{a}= T_{b}$. Under nonequilibrium conditions the curl flux $\mathcal{J}_{\rm curl}$ is generally nonvanishing, circulating in a directed loop among the three states. In Fig.~\ref{flux}(b), we plot the curl flux $\mathcal{J}_{\rm curl}$ as a function of $\Delta T$ for different $\lambda$. It can be seen that the curl flux increases monotonously with the temperature difference in the parameter regime considered. The nonequilibrium condition is thus manifested in the curl flux that drives the circulation dynamics among the populated states.

\subsection{Heat Current and Entropy Production}

In addition to the curl flux driving the circulation dynamics, the nonequilibrium nature of the system also leads to nonvanishing heat current and EPR (entropy production rate) at the steady state, reflecting steady-state transport features. To investigate the heat current associated with each reservoir in contact with the system, we rearrange the dissipators in the QME according to individual reservoir labels, so that we have $D[\rho]=D_0[\rho]+D_s[\rho]=D_a[\rho]+D_b[\rho]$. Here $D_a[\rho]$ and $D_b[\rho]$ represent the effect of the reservoirs in contact with cavity a and b, respectively. Their expressions are given by
\begin{subequations}\label{DRerservoirs}
\begin{equation}\label{Da}
\begin{split}
D_{a}[\rho]&=\frac{\gamma}{2}N_a^{A}[2A^{\dagger}\rho A-\rho AA^{\dagger}-AA^{\dagger}\rho]+\frac{\gamma}{2}N_a^{B}[2B^{\dagger}\rho B-\rho BB^{\dagger}-BB^{\dagger}\rho]\\
&+\frac{\gamma}{2}(N_a^{A}+1)[2A\rho A^{\dagger}-\rho A^{\dagger}A-A^{\dagger}A\rho]+\frac{\gamma}{2}(N_a^{B}+1)[2B\rho B^{\dagger}-\rho B^{\dagger}B-B^{\dagger}B\rho]\\
&+\frac{\gamma}{2}(N_a^{A}+N_a^{B})A^{\dagger}\rho B
-\frac{\gamma}{2}N_a^{A}BA^{\dagger}\rho-\frac{\gamma}{2}N_a^{B}\rho BA^{\dagger} +h.c.\\
&+\frac{\gamma}{2}(N_a^{A}+N_a^{B}+2)B\rho A^{\dagger}-\frac{\gamma}{2}(N_a^{A}+1)\rho A^{\dagger}B-\frac{\gamma}{2}(N_a^{B}+1)A^{\dagger}B\rho+h.c.,
\end{split}
\end{equation}
\begin{equation}\label{Db}
\begin{split}
D_{b}[\rho]&=\frac{\gamma}{2}N_b^{A}[2A^{\dagger}\rho A-\rho AA^{\dagger}-AA^{\dagger}\rho]+\frac{\gamma}{2}N_b^{B}[2B^{\dagger}\rho B-\rho BB^{\dagger}-BB^{\dagger}\rho]\\
&+\frac{\gamma}{2}(N_b^{A}+1)[2A\rho A^{\dagger}-\rho A^{\dagger}A-A^{\dagger}A\rho]+\frac{\gamma}{2}(N_b^{B}+1)[2B\rho B^{\dagger}-\rho B^{\dagger}B-B^{\dagger}B\rho]\\
&-\frac{\gamma}{2}(N_b^{A}+N_b^{B})A^{\dagger}\rho B
+\frac{\gamma}{2}N_b^{A}BA^{\dagger}\rho+\frac{\gamma}{2}N_b^{B}\rho BA^{\dagger}+h.c.\\
&-\frac{\gamma}{2}(N_b^{A}+N_b^{B}+2)B\rho A^{\dagger}+\frac{\gamma}{2}(N_b^{A}+1)\rho A^{\dagger}B+\frac{\gamma}{2}(N_b^{B}+1)A^{\dagger}B\rho+h.c..
\end{split}
\end{equation}
\end{subequations}
In the subspace spanned by  $\{|g\rangle, |e\rangle, |f\rangle\}$, the dissipators above have more explicit expressions, which will be discussed in the next section.

The heat current $J_i=\dot{Q}_i$ ($i=a,b$) that flows from the reservoir into the system at the steady state can then be derived as $J_i={\rm Tr}\{D_i[\rho_{ss}]H_s\}$~\cite{TW,KJ,JO,TW1,HT}. Direct calculation gives (see also next section)
\begin{subequations}\label{HeatCurrents}
\begin{eqnarray}
J_{a}&=&	\gamma\omega_{A}\left[N_{a}^{A}\rho_{gg}^{ss}
-(N_{a}^{A}+1)\rho_{ee}^{ss}-(N_{a}^{B}+1){\rm Re}(\rho^{\rm{coh}}_{ef})\right]\nonumber\\
	&+&\gamma\omega_{B}\left[N_{a}^{B}\rho_{gg}^{ss}
-(N_{a}^{B}+1)\rho_{ff}^{ss}-(N_{a}^{A}+1){\rm Re}(\rho^{\rm{coh}}_{ef})\right],\label{Ja}\\
J_{b}	&=&	\gamma\omega_{A}\left[N_{b}^{A}\rho_{gg}^{ss}
-(N_{b}^{A}+1)\rho_{ee}^{ss}+(N_{b}^{B}+1){\rm Re}(\rho^{\rm{coh}}_{ef})\right]\nonumber\\
	&+&\gamma\omega_{B}\left[N_{b}^{B}\rho_{gg}^{ss}
-(N_{b}^{B}+1)\rho_{ff}^{ss}+(N_{b}^{A}+1){\rm Re}(\rho^{\rm{coh}}_{ef})\right],\label{Jb}
\end{eqnarray}
\end{subequations}
where ${\rm Re}(\rho^{\rm{coh}}_{ef})$ represents the real part of the steady-state coherence. Notice that at the steady state the heat current flowing into the system from one reservoir should balance out the heat current flowing out of the system into the other reservoir in order to maintain the steady state. In other words, we have $J_a+J_b=0$, which can be verified from the expressions in Eq. (\ref{HeatCurrents}) and the steady-state QME [specifically, Eqs. (\ref{eq:ee}) and (\ref{eq:ff}) at the steady state].

One can also verify that at equilibrium with $T_a=T_b$, we have $J_a=J_b=0$, considering that $N_{a}^{A(B)}=N_{b}^{A(B)}=N^{A(B)}$, $\rho_{ee}^{eq}/\rho_{gg}^{eq}=N^A/(N^A+1)$, $\rho_{ff}^{eq}/\rho_{gg}^{eq}=N^B/(N^B+1)$, and $\rho^{\rm{coh}}_{ef}=0$ at equilibrium. In other words, the heat current vanishes at equilibrium as expected. At the nonequilibrium steady state with $T_a\neq T_b$, the heat current does not vanish; there is a continuous heat flow through the system from the high-temperature reservoir to the low-temperature reservoir. In Fig.~\ref{flux}(c), the heat current $J_b$ is plotted as a function of temperature difference $\Delta T$($=T_b-T_a>0$) for different $\lambda$. As one can see, the heat current increases as the temperature difference of the two reservoirs is increased.

At the steady state the EPR generated in the system balances out the rate of entropy flow out of the system, resulting in the expression ${\rm EPR}=-(J_b/T_b+J_a/T_a)$~\cite{QIT,jin1,jin2,jin3}. The negative sign in front comes from the fact that the heat currents are defined as those flowing into the system. Taking into account $J_a+J_b=0$, we have ${\rm EPR}=(1/T_a-1/T_b)J_b$, where $J_b$ is given in Eq. (\ref{Jb}). Obviously, EPR vanishes at equilibrium with $T_a=T_b$, indicating the reversible nature of the equilibrium steady state.  In Fig. \ref{flux}(d), EPR is plotted as a function of temperature difference for different $\lambda$. As can be seen, EPR increases with the temperature difference of the two reservoirs that characterizes the degree of nonequilibriumness. Such behaviors of EPR and heat current in relation to temperature difference can be well anticipated from a thermodynamic point of view.

From a physical perspective, detailed balance breaking indicating time irreversibility at the nonequilibrium steady state is reflected in the heat current flowing through the system and the nonvanishing EPR as a consequence of the temperature difference of the two reservoirs that is maintained at constant. Just like the power of a battery (arising from the electromotive force) eventually runs out, the temperature difference of reservoirs, in reality, also diminishes without maintenance. This leads to the energy cost in maintaining the nonequilibrium steady state of open quantum systems and its potentially beneficial properties such as steady-state quantum coherence and enhanced quantum metrology. To put it another way, energy supply and cost can be used to fight against environment-induced decoherence and deterioration of metrology, by maintaining a nonequilibrium steady state with quantum features that are robust in the interaction with the environments.

\section{Heat currents for population and coherence}
\label{explain}

A seemingly paradoxical result emerges when one examines how the various quantities in Fig.~\ref{coherencefisher} and Fig.~\ref{flux} behave in relation to the inter-cavity coupling strength $\lambda$. More specifically, with the increase of $\lambda$, the steady-state coherence (also the QFI in certain parameter regimes) decreases as seen in Fig.~\ref{coherencefisher}. In contrast, the nonequilibrium measures, including the curl flux, heat current and EPR, all increase with $\lambda$ shown in Fig.~\ref{flux}. Given that the steady-state coherence arises from the nonequilibrium condition, one would expect that it is tightly connected to the nonequilibrium measures (the curl flux, heat current and EPR). Thus the contrasting behaviors of the steady-state coherence and the nonequilibrium measures in relation to the inter-cavity coupling appear to be perplexing.

To unravel the mystery behind this seemingly counter-intuitive result, we track the heat current in and out of the system at a more detailed level. More specifically, we separate the heat current into a population component and a coherence component. To this end, we further divide the dissipator associated with each reservoir in Eq. (\ref{DRerservoirs}) into two parts, $D_i[\rho]=D_i^{(p)}[\rho]+D_i^{(c)}[\rho]\,(i=a,b)$, where $D_i^{(p)}[\rho]$ and $D_i^{(c)}[\rho]$ are the dissipators associated with population and coherence, respectively. In the subspace spanned by  $\{|g\rangle, |e\rangle, |f\rangle\}$, with $A=|g\rangle\langle e|$ and $B=|g\rangle\langle f|$, the expressions of $D_i^{(p)}[\rho]$ and $D_i^{(c)}[\rho]$ are given more explicitly by
\begin{subequations}
\begin{eqnarray}
D_{a}^{(p)}[\rho] & = & \frac{\gamma}{2}N_{a}^{A}\mathcal{L}_{1}[\rho]+\frac{\gamma}{2}N_{a}^{B}\mathcal{L}_{2}[\rho]+\frac{\gamma}{2}(N_{a}^{A}+1)\mathcal{L}_{3}[\rho]+\frac{\gamma}{2}(N_{a}^{B}+1)\mathcal{L}_{4}[\rho],\\
D_{a}^{(c)}[\rho] & = & \frac{\gamma}{2}(N_{a}^{A}+N_{a}^{B})\mathcal{L}_{5}[\rho]+\frac{\gamma}{2}(N_{a}^{A}+1)\mathcal{L}_{6}[\rho]+\frac{\gamma}{2}(N_{a}^{B}+1)\mathcal{L}_{7}[\rho],\\
D_{b}^{(p)}[\rho] & = & \frac{\gamma}{2}N_{b}^{A}\mathcal{L}_{1}[\rho]+\frac{\gamma}{2}N_{b}^{B}\mathcal{L}_{2}[\rho]+\frac{\gamma}{2}(N_{b}^{A}+1)\mathcal{L}_{3}[\rho]+\frac{\gamma}{2}(N_{b}^{B}+1)\mathcal{L}_{4}[\rho],\\
D_{b}^{(c)}[\rho] & = & -\frac{\gamma}{2}(N_{b}^{A}+N_{b}^{B})\mathcal{L}_{5}[\rho]-\frac{\gamma}{2}(N_{b}^{A}+1)\mathcal{L}_{6}[\rho]-\frac{\gamma}{2}(N_{b}^{B}+1)\mathcal{L}_{7}[\rho].
\end{eqnarray}
\end{subequations}
where
\begin{subequations}
\begin{eqnarray}
\mathcal{L}_{1}[\rho] & = & 2|e\rangle\langle g|\rho|g\rangle\langle e|-\rho|g\rangle\langle g|-|g\rangle\langle g|\rho,\\
\mathcal{L}_{2}[\rho] & = & 2|f\rangle\langle g|\rho|g\rangle\langle f|-\rho|g\rangle\langle g|-|g\rangle\langle g|\rho,\\
\mathcal{L}_{3}[\rho] & = & 2|g\rangle\langle e|\rho|e\rangle\langle g|-\rho|e\rangle\langle e|-|e\rangle\langle e|\rho,\\
\mathcal{L}_{4}[\rho] & = & 2|g\rangle\langle f|\rho|f\rangle\langle g|-\rho|f\rangle\langle f|-|f\rangle\langle f|\rho,\\
\mathcal{L}_{5}[\rho] & = & |e\rangle\langle g|\rho|g\rangle\langle f|+|f\rangle\langle g|\rho|g\rangle\langle e|,\\
\mathcal{L}_{6}[\rho] & = & |g\rangle\langle f|\rho|e\rangle\langle g|+|g\rangle\langle e|\rho|f\rangle\langle g|-\rho|e\rangle\langle f|-|f\rangle\langle e|\rho,\\
\mathcal{L}_{7}[\rho] & = & |g\rangle\langle f|\rho|e\rangle\langle g|+|g\rangle\langle e|\rho|f\rangle\langle g|-\rho|f\rangle\langle e|-|e\rangle\langle f|\rho.
\end{eqnarray}
\end{subequations}

\begin{figure}[tbp]
\centering
\includegraphics[width=1.0\columnwidth]{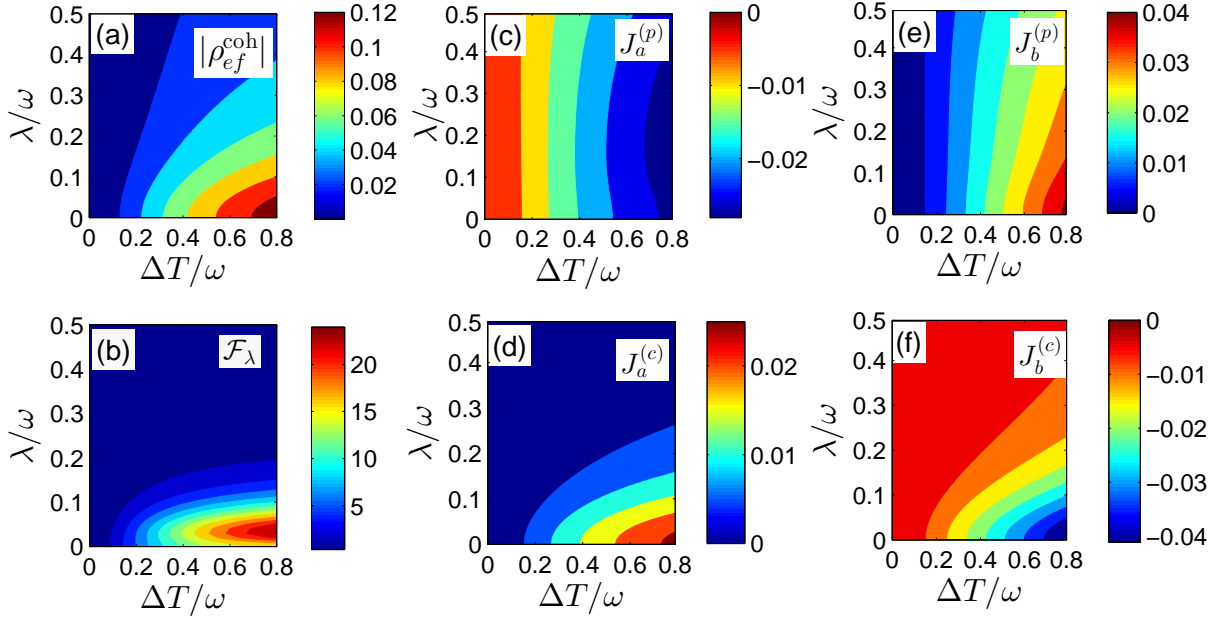}
\caption{(Color online) The steady-state coherence, QFI, and heat currents as functions of inter-cavity coupling $\lambda$ and temperature difference $\Delta T=T_b-T_a$. The parameters are set as $\gamma=0.1\omega$, $T_a=0.2\omega$.}
\label{physics}
\end{figure}

Accordingly, the heat currents can be defined more specifically as $J_i^{(n)}={\rm Tr}\{D_i^n[\rho_{ss}]H_s\}$ where $i=a,b$ and $n=p,c$, for each reservoir and for population and coherence components individually. Direct calculation yields
\begin{subequations}\label{Jpc}
\begin{eqnarray}
J_{a}^{(p)}	&=&	\gamma\left(\omega_{A}N_{a}^{A}+\omega_{B}N_{a}^{B}\right)\rho_{gg}^{ss}
-\gamma\omega_{A}(N_{a}^{A}+1)\rho_{ee}^{ss}-\gamma\omega_{B}(N_{a}^{B}+1)\rho_{ff}^{ss},\label{Jap}\\
J_{b}^{(p)}	&=&	\gamma\left(\omega_{A}N_{b}^{A}+\omega_{B}N_{b}^{B}\right)\rho_{gg}^{ss}
-\gamma\omega_{A}(N_{b}^{A}+1)\rho_{ee}^{ss}-\gamma\omega_{B}(N_{b}^{B}+1)\rho_{ff}^{ss},\label{Jbp}\\
J_{a}^{(c)}	&=&	-\gamma\left[ \omega_{A}(N_{a}^{B}+1)+\omega_{B}(N_{a}^{A}+1)\right] {\rm Re}(\rho^{\rm{coh}}_{ef}), \label{Jac}\\
J_{b}^{(c)}	&=&	\gamma\left[ \omega_{A}(N_{b}^{B}+1)+\omega_{B}(N_{b}^{A}+1)\right] {\rm Re}(\rho^{\rm{coh}}_{ef}).\label{Jbc}
\end{eqnarray}
\end{subequations}

It can be verified that $J_{i}=J_{i}^{(p)}+J_{i}^{(c)}$ ($i=a, b$), which means the heat current associated with each reservoir has been decomposed into a population component and a coherence component. The population heat currents $J_a^{(p)}$ and $J_b^{(p)}$, dependent only on the steady-state populations, are associated with maintaining populations at the nonequilibrium steady state away from their equilibrium values.  On the other hand, the coherence heat currents $J_a^{(c)}$ and $J_b^{(c)}$, directly related to the steady-state coherence, are associated with maintaining nonvanishing quantum coherence at the nonequilibrium steady state. It is only this part of the heat current that is tightly connected to the steady-state coherence.

In Fig.~\ref{physics}, we contour plot the steady-state coherence (also the closely related QFI) and the heat currents as functions of temperature difference $\Delta T$ and inter-cavity coupling strength $\lambda$. From this figure one can see that $J_a^{(p)}<0$ and $J_b^{(p)}>0$ (with $T_a<T_b$), which means the system absorbs energy from the high-temperature reservoir and discharges it into the low-temperature reservoir, in maintaining the nonequilibrium steady-state populations. On the other hand, it can also be seen that $J_a^{(c)}>0$ and $J_b^{(c)}<0$, which means, in maintaining nonvanishing steady-state coherence, the system absorbs energy from the low-temperature reservoir and releases it into the high-temperature reservoir. In other words, maintaining the steady-state coherence forms an inverse heat current. However, the total heat current associated with each reservoir, with the population and coherence heat currents combined together, still points from the high-temperature reservoir,  through the system, to the low-temperature reservoir, ensuring a positive EPR in agreement with the second law of thermodynamics.

Moreover, one can see clearly that the coherence heat current $J_a^{(c)}$  (or consider $-J_b^{(c)}>0$) displays a similar pattern to that of the steady-state coherence $|\rho^{\rm{coh}}_{ef}|$ in relation to $\Delta T$ and $\lambda$. This confirms yet again the perspective derived from the expressions in Eq. (\ref{Jpc}) that only a part of the heat current, namely the coherence heat current, rather than the total heat current (including the population heat current), is closely related to the steady-state coherence (and thus enhanced quantum metrology in the weak inter-cavity coupling regime). Therefore, the paradox mentioned in the beginning of this section is resolved by realizing that the steady-state coherence is not so tightly connected to the total nonequilibrium measures but only to a part of them.

\section{Remarks and Conclusion}
\label{conclusion}

There is a possibility that our predictions on the coherence enhanced quantum metrology may be experimentally tested in the foreseeable future. Our scheme of optical molecule can be realized in coupled superconducting transmission line cavities, which support single-mode electromagnetical field with resonant frequency $\omega/2\pi\approx3$GHz~\cite{AW}. To construct nonequilibrium reservoirs at a temperature difference in the range of tens of mK, one can separate the two cavities in a distance of several cm and control their bath temperatures by diluted magnetic refrigerators. The nonequilibrium condition can also be realized by coupling cavities to reservoirs with different coupling strengths.  Via tunable capacitances, the inter-cavity coupling strength in the range $\lambda/2\pi \approx 5-100$MHz can be achieved ~\cite{AW,MM}. The heat current in this kind of solid-state systems can be quantified indirectly by measuring the thermal conductance with the assistance of scanning electron microscope imaging~\cite{CWC}. The steady-state coherence can be measured using two dimensional spectroscopy, where three laser pulses interact in the weak field limit with the sample to produce a third-order polarization, and the cross peak in the two dimensional spectroscopy quantifies quantum coherence~\cite{GPD,TT,mukamel}. Moreover, it has been recently proposed that, by comparing the measurement statistics of a state before and after a small unitary rotation, the lower bounds on the QFI can be determined~\cite{mQFI}. Therefore, there is a good chance the theoretical and numerical results presented in this paper can be compared with experiments in the near future.

In summary, in this work we investigated the effect of coherence enhanced quantum metrology in an optical molecular system interacting with nonequilibrium environments. The model we considered consists of a pair of coupled single-mode optical cavities, each in contact with its own reservoir at a different temperature. We studied this model both analytically and numerically, based on the QME beyond secular approximation. We obtained the analytical solution to the steady state of the QME, and found that there is nonvanishing steady-state quantum coherence, which is sustained by the nonequilibrium condition characterized by the temperature difference of the two reservoirs. We showed that the steady-state coherence makes a positive contribution to the QFI in addition to its classical part, and that quantum metrology quantified by QFI can be effectively enhanced by the steady-state coherence in the weak inter-cavity coupling regime. To quantify the measures of the nonequilibrium condition on both the dynamical level and the thermodynamic level, we investigated the curl flux driving the circulation dynamics as well as the heat current and EPR associated with the thermal transport process, in relation to the temperature difference characterizing the degree of nonequilibriumness. A seemingly paradoxical feature emerged that these nonequilibrium measures displayed a contrastingly different trend  from that of the steady-state coherence (and QFI in certain parameter regimes) in relation to the inter-cavity coupling strength. By decomposing the heat current into two parts associated with maintaining the steady-state population and coherence, respectively, we resolved the paradox by showing that the steady-state coherence is tightly tied to only a part of the nonequilibrium measures. In addition, we had an interesting discovery that the heat current associated with maintaining the steady-state coherence flows from the low-temperature reservoir to the high-temperature reservoir, but this process is not in violation of the second law of thermodynamics when the population heat current is also taken into account.

Our work provides a viable way to enhance quantum metrology with improved precision of parameter estimation for open quantum systems, by exploiting the stable quantum coherence at the nonequilibrium steady state, at the cost of energy supply to maintain the nonequilibrium condition. The nonequilibrium condition can be controlled and manipulated to maximize the utility of the steady-state coherence in quantum metrology. The potential applications of our work is not limited to the field of quantum metrology, but also extend to quantum technology in general, including device designing and quantum computation.

%nonequilibriumness enhances the  we have studied how quantum metrology through inducing the steady-state coherence. We illustrate this with an example of

\appendix

\section{Expressions of $\mathcal{N}$ and $\mathcal{A}$}\label{Expressions}

The normalization factor $\mathcal{N}$ in the analytical solution of the steady state in Eqs. (\ref{Population2}) and (\ref{SSCES2}), fixed by the condition $\rho_{gg}^{ss}+\rho_{ee}^{ss}+\rho_{ff}^{ss}=1$, has the expression
\begin{equation}\label{ExpN}
\begin{split}
\mathcal{N}&=\left[(N_{+}^{A}+1)(N_{+}^{B}+1)-(N_{+}^{A}+N_{+}^{B}+2)N_{-}^{A}N_{-}^{B}R\right]\\
&+\left[N_{+}^{A}(N_{+}^{B}+1)-(N_{+}^{A}+N_{+}^{B}+1)N_{-}^{A}N_{-}^{B}R-(N_{-}^{B})^2R\right]\\
&+\left[N_{+}^{B}(N_{+}^{A}+1)-(N_{+}^{A}+N_{+}^{B}+1)N_{-}^{A}N_{-}^{B}R-(N_{-}^{A})^2R\right],
\end{split}
\end{equation}
where $R$ is defined in Eq. (\ref{ExpR}).

The elements of the $3\times 3$ matrix $\mathcal{A}$, defined as $\mathcal{A}=\mathcal{M}_p-\mathcal{M}_{pc}\mathcal{M}_c^{-1}\mathcal{M}_{cp}$, read as follows:
\begin{equation}\label{ExpA}
\begin{split}
\mathcal{A}_{gg}&=2\gamma\left[-(N_+^A+N_+^B)+(N_-^A+N_-^B)(N_-^A+N_-^B)R\right]\\
\mathcal{A}_{ee}&=2\gamma\left[-(N_+^A+1)+ N_-^AN_-^B R \right]\\
\mathcal{A}_{ff}&=2\gamma\left[ - (N_+^B+1)+ N_-^AN_-^B R\right]\\
\mathcal{A}_{ge}&=2\gamma\left[(N_+^A+1) -(N_-^A+N_-^B)N_-^A R\right]\\
\mathcal{A}_{gf}&=2\gamma\left[(N_+^B+1) -(N_-^A+N_-^B)N_-^B R \right]\\
\mathcal{A}_{eg}&=2\gamma\left[ N_+^A-(N_-^A+N_-^B)N_-^BR\right]\\
\mathcal{A}_{fg}&=2\gamma\left[ N_+^B - (N_-^A+N_-^B)N_-^AR\right]\\
\mathcal{A}_{ef}&=2\gamma (N_-^B)^2 R \\
\mathcal{A}_{fe}&=2\gamma(N_-^A)^2 R\\
\end{split}.
\end{equation}

\section{Method of obtaining the analytical steady-state solution}\label{SolutionMethod}

One may try to obtain the steady-state solution by solving directly the set of linear algebraic equations from Eq. (\ref{elements}) at the steady state (the last two equations can be excluded as they are decoupled from the rest). But the results are too complicated in form and without clear meaning. Instead, we obtain the analytical solution using the technique of `dimension reduction', which makes the solution more manageable.

We notice that  the steady-state coherence can be expressed in terms of the steady-state populations. More specifically, at the steady state Eq. (\ref{eq:offd}) leads to
\begin{equation}\label{SSCForm}
\rho_{ef}^{\rm coh}=\frac{(N_{-}^{A}+N_{-}^{B})\rho_{gg}^{ss}- N_{-}^{A}\rho_{ee}^{ss}- N_{-}^{B}\rho_{ff}^{ss}}{(N_{+}^{A}+N_{+}^{B}+2)+i(\omega_{A}-\omega_{B})/\gamma}.
\end{equation}
Thus we only need to solve the steady-state populations, which are determined by $\mathcal{A}\rho_p^{ss}=0$ [see Eq. (\ref{landscale})], where $\mathcal{A}$ is a $3\times 3$ matrix with its elements given in Eq. (\ref{ExpA}).

It is easy to check that each column of $\mathcal{A}$ adds up to zero (a property associated with probability conservation), indicating that its determinant is zero. For a generic $3\times 3$ matrix with such a property, it can be directly verified that $\rho_p^{ss}$ with the following form
\begin{equation}\label{SSPForm}
\rho_p^{ss}=\frac{1}{\mathcal{N}}
\begin{bmatrix}
\mathcal{A}_{22}\mathcal{A}_{33}-\mathcal{A}_{23}\mathcal{A}_{32}\\
\mathcal{A}_{31}\mathcal{A}_{23}-\mathcal{A}_{21}\mathcal{A}_{33}\\
\mathcal{A}_{21}\mathcal{A}_{32}-\mathcal{A}_{31}\mathcal{A}_{22}\\
\end{bmatrix}
\end{equation}
satisfies $\mathcal{A}\rho_p^{ss}=0$. In the above, $\mathcal{N}$ is a normalization factor and the matrix element subscripts $1$, $2$, $3$ correspond to $g$, $e$, $f$, respectively, in our particular system. Typically, physical conditions ensure that the steady state is unique up to normalization (mathematically this means the rank of $\mathcal{A}$ is $2$); then $\rho_p^{ss}$ above will be the only steady-state solution, up to normalization.

Inserting the expressions of the matrix elements  in Eq. (\ref{ExpA}) into Eq. (\ref{SSPForm}) and fixing the factor $\mathcal{N}$ by the probability normalization condition $\rho_{gg}^{ss}+\rho_{ee}^{ss}+\rho_{ff}^{ss}=1$, we obtain the steady-state populations. Then the steady-state coherence is calculated according to Eq. (\ref{SSCForm}). Eventually, we reach the steady-state solution given in  Eqs. (\ref{Population2}) and (\ref{SSCES2}).

The above approach can be extended to more general scenarios. Consider the QME in the vector-matrix form $\mathcal{M}|\rho_{ss}\rangle=0$. The steady-state coherence can be expressed in terms of the steady-state population by $\rho_c^{ss}=-\mathcal{M}_{c}^{-1}\mathcal{M}_{cp}\rho_p^{ss}$, resulting from the coherence component of Eq. (\ref{LSQME}) at the steady state. The steady-state populations are determined by the equation $\mathcal{A}\rho_p^{ss}=0$ [Eq. (\ref{landscale})] with reduced dimension. Assuming that the solution $\rho_p^{ss}$ is unique up to normalization (i.e., the rank of $\mathcal{A}$ is $n-1$), $\rho_p^{ss}$ can be obtained as follows. Choose any row of $\mathcal{A}$, say the first row, with the elements $(\mathcal{A}_{11}, \cdots, \mathcal{A}_{1i},\cdots, \mathcal{A}_{1n} )$. Then the $i$-th component of $\rho_p^{ss}$ is proportional to the cofactor (signed minor) of $\mathcal{A}_{1i}$. The form in Eq. (\ref{SSPForm}) is an example of this rule. After obtaining $\rho_p^{ss}$ one can then calculate $\rho_c^{ss}$, thus obtaining the full steady-state solution. As the dimension increases, however, analytical solutions quickly become impractical even with this dimension reduction technique.

\begin{acknowledgments}
We thank X. X. Yi for his helpful discussions. This work is supported by the NSFC (under Grant No.11404021), the Jilin province science and technology development plan item (under Grant No. 20170520132JH) and the Fundamental Research Funds for the Central Universities (under Grant Nos. 2412016KJ015 and 2412016KJ004). GC and JW thank the support in part from NSF PHY 76066.
\end{acknowledgments}

\end{document}